\documentclass[12pt]{iopart}
\usepackage{graphicx} 
\usepackage{epstopdf}
\usepackage{psfrag}
\usepackage{epsfig}
\usepackage{subfigure} 

\begin{document}

\title[Effectiveness of a Dynein team in tug of war]{Effectiveness of a dynein team
  in tug-of-war helped by reduced load-sensitivity of detachment: 
evidence from study of bidirectional endosome
  transport in {\it D. discoideum}}

\author{Deepak Bhat, Manoj Gopalakrishnan}

\address{Department of Physics, Indian Institute of Technology Madras,
  Chennai 600036, India}

\ead{manoj@physics.iitm.ac.in,deepak@physics.iitm.ac.in}

\date{\today}

\begin{abstract}
 Bidirectional cargo transport by molecular motors in cells is a
 complex phenomenon, in which the cargo (usually a
 vesicle) alternately moves in retrograde and anterograde directions.
 In this case, teams of oppositely pulling motors (eg., kinesin and
 dynein) bind to the cargo simultaneously, and  `coordinate' their
 activity such that the motion consists of spells of positively and
 negatively directed segments, separated by pauses of varying
 duration. A set of recent experiments have analyzed the bidirectional
 motion of endosomes in the amoeba {\it D. discoideum} in detail. It
 was found that in between directional switches, a team of 5-6 dyneins
 stall a cargo against a stronger kinesin in tug of war, which lasts for almost a
 second. As the mean detachment time of a kinesin
 under its stall load was also observed to be $\sim$ 1s, we infer that  the collective
 detachment time of the dynein assembly must also be similar.                                                  
 Here, we analyze this inference from a modeling perspective, using
 experimentally measured single-molecule parameters as inputs. We find
 that the commonly assumed exponential load-dependent detachment rate
 is inconsistent with observations, as it predicts that a 
5-dynein assembly will detach under its combined stall load in less
than a hundredth of a second. A modified model where the load-dependent unbinding rate is assumed to saturate at
 stall-force level  for super-stall loads gives results which are in
 agreement with experimental data. Our analysis suggests that the
 load-dependent detachment of a dynein in a team is qualitatively
 different at sub-stall and super-stall loads, a conclusion which is
 likely to have implications in other situations involving collective
 effects of many motors.
\end{abstract}  

\submitto{\PB}

\section{Introduction:}

Active cellular transport is made possible by ATP-consuming
motor-proteins, i.e., kinesin, dynein and myosin which walk on the
polar cytoskeletal filaments.  Kinesin and Dynein are
microtubule-associated motors while myosin is actin-based. The
directionality of motor motion is derived from the structural polarity
of the corresponding filament and the irreversibility of ATP
hydrolysis reaction which powers the motion. Dynein is involved in
minus directed (towards center of the cell) transport where as, kinesin
shows plus (towards periphery of the cell) directed motion on
microtubules.  Both kinesin and dynein are involved in transport of
intracellular vesicles and other cargoes.  Many intracellular cargoes are known to move in a
bidirectional fashion, where the direction of motion is reversed
typically every few seconds (reviewed in \cite{GROSS,WELTE}). Examples of such  cargoes include
mitochondria, pigment granules, endosomes and viruses. Bidirectional
motion results from competition between oppositely pulling motors,
which bind the same cargo. Two mechanisms have been envisaged in this
context, both of which can cause bidirectional motion, i.e., (a) {\it
  tug of war} and (b) {\it reciprocal coordination}. In the first
case, motors actively pull against each other, leading to 
tug of war (TOW) situations. When one team wins (by enhancing detachment of
the opposite team), it moves the cargo in its direction, until one or
more of the other motors bind. In the second case, a pre-existing
regulatory protein complex is assumed to mediate the motor-filament
interaction, which switches on and switches off the plus and
minus-directed motors alternately. Several candidates have been
proposed to perform the role of such a complex, eg. Klar and dynactin\cite{WELTE}.
 
 Recently, Soppina et al \cite{SOPPINA} reported a series of in vivo and in vitro
 experiments with endosomes in the amoeba {\it Dictyostelium
   discoideum} (henceforth, simply {\it Dictyostelium}). The
 observed endosome motion was bidirectional, and showed three
 distinct phases, including positively and negatively directed
 unidirectional motion and a TOW phase where the cargo stalled
 as a result of being pulled in both directions by motor teams
 exerting approximately
 equal force (although a small negatively directed velocity of a few
 nm/s was observed in this phase, indicating that one motor team was
 slightly stronger than the other). The endosome was also observed to
 undergo fission in some of the TOW situations, which suggests
 a new functional role for the bidirectional transport. 

 Detailed analysis using optical trapping suggested that an endosome is
 transported by one kinesin and 5-6 cytoplasmic dyneins. It was determined that the stall force of the kinesin
 is 5.5 pN, while that for dynein is 1.1 pN, showing that the kinesin
 is almost 5 times stronger than a single dynein. Furthermore, the
 motors were found to have very different tenacity: under stall load, it
 was determined that the kinesin would
 detach in a time scale of about 1 second. Since the mean duration of the TOW
 was also close to 1 second, it is necessary that the opposing team of
 dyneins do not detach faster (as then the TOW would end 
 sooner, and preferentially in favor of kinesin). The mean detachment
 time of a single dynein under an opposing load equal to its  stall
 force was determined to be only about 0.2 seconds. Therefore, it is
 important to ensure that, as the number of dyneins in the team is
 increased, the mean detachment time under the (combined) stall load (later referred to as the {\it stall time}) 
 also increases, so that a team of 5-6 dyneins is an effective competitor
 against a strong, tenacious kinesin. The relationship between the stall time of a motor assembly and the number of motors 
 is far from obvious, though. As we show in a later section, in the context of the present experiments, increase of stall time 
 with motor number requires that the detachment rate, which normally increases under
 opposing load, saturates (or possibly increases much slower than
 exponential) at super-stall loads.  We
 will show that this conjecture is consistent with the endosomal
 bidirectional transport data reported in\cite{SOPPINA} and the human adenovirus
 transport data reported in\cite{ADENO}. 
 
From a more general perspective, to what extent the properties of a
multiple-motor assembly may be deduced from the physical parameters
that characterize the individual motors, is a question that merits
attention. Recent experimental observations have thrown up many
surprises, such as a cargo carried by two motors does not necessarily
travel longer on a filament without detachment compared to one
motor {\it in vivo}\cite{SHUBEITA}. Computational modeling of multiple motor systems
with built-in interaction between motors has predicted sub-additive
stall forces\cite{KUNWAR}. These results indicate that
the different motors working in a team likely interact with each
other. However, such interactions are difficult to model directly
owing to the complex structure of motor proteins. Therefore, an indirect way to
deduce information about motor-motor interactions would be to compare
the predictions of simple {\it in silico} motor models (where direct interactions are
deemed absent) with experimental results. 

A recent interesting paper by Kunwar et. al.\cite{KUNWAR1} has reported experimental and modeling studies of 
bidirectional transport of lipid droplets in {\it Drosophila} embryos. In this work, the authors attempted to fit the experimentally 
observed parameters of transport, like the mean pause duration, frequency of pauses and the statistics of the unidirectional run lengths 
to predictions from a tug-of-war model. The authors used both a `mean-field' TOW model (as in,  eg., \cite{MULLER, MJMULLER}) where same-polarity 
motors share opposing load equally, as well as a more sophisticated model where load-sharing is stochastic and uneven\cite{KUNWAR}. 
It is reported that neither of these models agree quantitatively 
with the in vivo experiments, suggesting that further regulatory mechanisms may be in place to regulate the transport. Interestingly, {\it in vitro} bead 
experiments also showed that the detachment rate of dynein as a function of opposing force is non-monotonic, with a peak around 2pN and possibly saturating 
at very high loads (up to 10pN), in disagreement with the commonly used exponential detachment rate. It was also noted that the disagreement with 
experimental data would be worse if the exponential detachment rate is used, instead of the non-monotonic, experimentally determined curve. 

In this paper, we report a detailed analysis of the bidirectional endosome transport data reported in Soppina et. al. The individual single molecule parameters are 
extracted from in vitro data reported in the same paper. These parameters are then used to simulate the bidirectional endosome transport {\it in silico}, using the stochastic model first introduced in M\"uller et. al.\cite{MULLER,MJMULLER}, which itself is based on the earlier work by Klumpp and Lipowsky\cite{KLUMPP}. We show that the the observed pause/TOW durations are inconsistent with an exponential increase in detachment rate with opposing load, for dyneins. A modified detachment rate which saturates above the stall force is found to agree much better with the experimental data,  which is consistent with (and arrived independently of) the lipid droplet results presented in \cite{KUNWAR1}. Moreover, similar to Kunwar et. al., we are also unable to reproduce the experimentally observed 
unidirectional  run lengths within our model, suggesting, among other possibilities, the presence of a regulating factor/switch. 
 
 In the rest of the paper, we first describe the experimental results
 of Soppina et. al. in brief. We then explain the salient features of
 our model, with the assumptions used, analyze the experimental
 results using the model, discuss our results and present our conclusions.
 
 \section{Review of experimental results for endosome transport in Dictyostelium}

Several experiments have shown bidirectional motion of cargoes inside
living cells, eg., pigment granules\cite{GELFAND}, lipid
droplets\cite{GROSS1,VALETTI}, endosomes\cite{VALETTI,MURRAY} and
several types of viruses\cite{SMITH,SUOM,MCDONALD}. However, in most
cases, the precise mechanism that regulates the switching between plus
and minus end directed motion remains unclear. In contrast, a few 
experiments\cite{SOPPINA,GENNERICH} have shown
direct evidence for TOW between opposing teams of motors. In
Soppina et. al.\cite{SOPPINA}, the motion of early endosomes in the
amoeba {\it Dictyostelium} was studied in detail, and parameters
characterizing the bidirectional transport are measured. {\it In vivo}
observations showed bidirectional motion of the cargo with frequent
`triphasic' reversals, which occasionally terminated in fission of
the endosomes. Reversal in direction was characterized by a slow phase 
(cargo velocity of a few nm/s) sandwiched between two oppositely
directed, fast unidirectional runs (velocity nearly 2.1 $\mu$m/s). The
mean duration of the slow phase was about 0.8 seconds. Endosomes elongated 
up to 25\% of its length in this slow moving phase which sometimes ended in
fission. Forces of 11-18pN are sufficient to produce tubes from  
membrane of golgi apparatus and endoplasmic reticulum \cite{ARPITA}.
So, the elongation and fission of endosome is because of high activity of one to two 
kinesins pulling against five to eight dyneins which can effectively produce 
tension of 11-18pN. 

Soppina et. al. also performed a number of {\it in vitro} optical trap
experiments to complement the {\it in vivo} observations. Endosome
motion was reconstituted in {\it Dictyostelium} cell extracts, and the
motion was very similar to that observed {\it in vivo}. Individual
molecular motor properties were analyzed using optical trap
on motor coated plastic beads. Molecular motors purified from {\it
  Dictyostelium}  cells were bound to the beads and made to
move under controlled loads. Single motor run lengths and velocities were
measured for both dynein and  kinesin.  Stall force for one dynein is
found to be 1.1pN whereas that for kinesin is 5.5pN. Stall force 
for two and three dyneins were 2.2pN  and 3.3pN respectively. 
The additive stall force for multiple motors showed equal
load sharing by engaged motors. Importantly, the
mean duration  for which a single motor or a team of motors survive in the trap
without detachment against its respective stall force (henceforth
referred to as ``stall time'') was measured using beads coated
with varying number of dyneins. 

Table \ref{tab:tab1} summarizes the results from
{\it in vitro} studies. Note that the stall time  increases almost linearly with the
number of dyneins, and extrapolation to five dyneins would give a stall time of 1s, 
which matches closely with the observed TOW duration {\it in vivo}. The endosomes 
also undergo considerable stretching during the TOW events as discussed previously, and the time-scale 
of elongation may vary between a fraction of a second to  to almost a second (Fig.S2, Supplementary Index of \cite{SOPPINA}). 
This time-delay in the transmission of load across the endosome caused by elongation is also likely to be important in determining the TOW duration {\it in vivo}, but this obviously does not apply to motor-coated beads which show clearly an upward trend in the stall time with the number of engaged motors. Therefore, in this paper, we will focus on cargo-independent motor properties as far as explaining the stall time or TOW duration is concerned. 

A summary of important parameters characterizing {\it in vivo} observations of bidirectional endosome transport is given in
Table \ref{tab:tab2}. In the next section, we discuss our mathematical model which is used
to analyze these observations.

\begin{table}
\begin{tabular}{|c|c|c|c|c|c|}
\hline Kind of motor&motor number ($n$)& velocity $v_{\pm}$ &runlength $l_{\pm}$ &Stall force($F_s$) & Stall time\\ 
& & ($\mu m s^{-1}$) & ($\mu$m) & (pN) &  $T_{\pm}^n$($F_{\pm}^s$) ~~(s)\\\hline 
            &      1&     1.2&      1.8&     1.1&      0.2$\pm$ 0.1\\  
Dynein (--) &      2&     1.5&        8&     2.2&      0.4$\pm$  0.2\\  
            &      3&     1.5&       15&     3.3&      0.6 $\pm$ 0.4\\\hline
Kinesin (+) &      1&     1.6&      5.1&     5.5&      1.1 \\  
            &      2&     1.6&       15&      11&      ND \\ \hline
\end{tabular}
\caption{ A list of parameters measured from {\it in vitro} motion of
  motor-coated plastic beads, as reported in \cite{SOPPINA}. }
\label{tab:tab1}
\end{table}

\begin{table}
\begin{tabular}{|c|c|c|c|}
\hline
Type of event & pre-reversal run time(s) & TOW duration(s) & post-reversal run time(s)\\
\hline
plus$\rightarrow$ minus & 5 & 0.9 & 8.69\\\hline
minus$\rightarrow$ plus & 8.69 & 1.1 & 5\\
\hline 
\end{tabular}
\caption{The table lists the bidirectional transport parameters
  obtained from motion of endosomes in {\it Dictystelium} cell
  extracts, as reported in \cite{SOPPINA}, Table S1 therein. All times
  are in seconds. The run times are obtained from run lengths (measured)
  and velocities (measured to be in the range $\sim$ 2.0-2.3$\mu$m/s). Error bars are
  omitted from this table.}
\label{tab:tab2}
\end{table}

\section{Stochastic model}

For quantitative analysis of the data, we use the stochastic model described in M\"uller et
al \cite{MULLER}. In this model, a cargo is assumed to be bound to a certain fixed
number of motors  of either directionality. The motors are assumed to
be always bound to the cargo, but may attach to and detach from the filament stochastically. The opposing load is assumed to be shared by similar motors equally, which is a simplifying feature of this `mean-field' model, in which there is no direct motor-cargo or motor-motor interactions.  Other models have been proposed  which include both motor-cargo\cite{KUNWAR,NEW2} and motor-motor\cite{BOUZAT1,BOUZAT2} interactions in more detail, as a consequence of which load-sharing by motors also becomes stochastic. In this paper, we ignore these effects for the sake of simplicity. 

Consider a cargo with $D$ attached dyneins and $K$ attached kinesins moving on a microtubule. 
The state of the cargo at a given point of time  may be denoted by $|{dk}\rangle$, where 
$d\leq D$ is the number of dyneins attached to the filament, and $k\leq K$ is the number of
kinesins attached to the filament. In the model in \cite{MULLER}, the forces exerted by individual motors are determined dynamically 
through a set of self-consistent equations based on balance of forces on the cargo, and  phenomenological equations that describe 
the decrease in velocity with load. These equations yield the cargo velocity $v_c(d,k)$ and the dynamic load $F_c(d,k)$ as functions 
of the numbers of filament-engaged motors. While calculating $v_c(d,k)$ and $F_c(d,k)$, a piecewise linear force-velocity relation is assumed for each type of motor:

 \begin{eqnarray}
v_{\pm}(\lambda_{\pm})= v_{\pm}\left(1-\frac{\lambda_{\pm}}{F^{\pm}_s}\right)~~~~~~~~~~~:  \lambda_{\pm}\leq F^{\pm}_S\nonumber\\~~~~~~~= v^{\prime}_{\pm}\left(1-\frac{\lambda_{\pm}}{F^{\pm}_s}\right)
~~~~~~~~~~~:  \lambda_{\pm} \geq F^{\pm}_S
\label{eq:eq0}
\end{eqnarray}

where $v_{\pm}$ and $v^{\prime}_{\pm}$ are absolute values of motor forward and backward velocities, while $\lambda_{\pm}$ denote the load per motor (under the assumption of equal load sharing), given by  

\begin{equation}
\lambda_{+}=\frac{F_c(d,k)}{k}~~~;~~~\lambda_{-}=\frac{F_c(d,k)}{d}
\label{eq:eq0+}
\end{equation}

For the sake of consistency in the model, a non-zero absolute backward velocity $v^{\prime}_{\pm}$ is considered for both the motors with values $v^{\prime}_+$=6 nm s$^{-1}$ for  kinesin and $v^{\prime}_{-}$=72nm s$^{-1}$ for dynein \cite{MULLER}. Forward velocity of a kinesin and a dynein respectively taken as $v_+=1.6 \mu$m s$^{-1}$ and $v_-=1.2 \mu$m s$^{-1}$ from in-vitro studies reported in \cite{SOPPINA}(see Table \ref{tab:tab2}). We would like to emphasize that the precise form of the force-velocity curve and the absolute values of of the backward velocities are not crucial ingredients for the model.

Let $\epsilon_{+}(d,k)$ and $\pi_{+}$ denote the detachment 
and attachment rate of an individual kinesin, and $\epsilon_{-}(d,k)$ and $\pi_{-}$ denote the 
same for an individual dynein, when the cargo is in a state $|{dk}\rangle$. Like forward and backward velocities, 
the detachment rates also depend on the opposing load per motor:

\begin{equation}
\epsilon_{\pm}=\epsilon_{\pm}^{0}f_{\pm}(\lambda_{\pm})
\label{eq:eq1}
\end{equation}

where the functions $f_{\pm}(\lambda)$ specify the dependence of the detachment rates on the 
load per motor $\lambda_{\pm}$ (given by Eq.\ref{eq:eq0+}), and $\epsilon_{\pm}^{0}$ are the {\it intrinsic} detachment rate of 
each motor, in the absence of any opposing load (by construction, we then require $f_{\pm}(0)=1$). 
The precise nature of the function $f_{+}(\lambda)$ has been determined in optical trap experiments 
for kinesin\cite{COPPIN}. It is seen that the dissociation rate (and hence $f_{+}$) increases 
rapidly (almost linearly) with load, slowing down close to the stall load ($\approx 5$pN). For
modeling purposes, it is commonly assumed (based on Kramer's rate theory) that $f_{\pm}(\lambda)$ 
depends exponentially on $\lambda$, i.e., 

\begin{equation}
f_{\pm}(\lambda)=\exp\left(\frac{\lambda}{f_{\pm}^{d}}\right),
\label{eq:eq2}
\end{equation}

where the parameters $f_{\pm}^{d}$ are specific to each type of
motors, and are called {\it detachment forces}. We will see later
that this exponential form in Eq.\ref{eq:eq2}, though widely used in modeling 
studies\cite{MULLER,MJMULLER,KLUMPP,JULICH,FISHER}, turns out  to be unsuccessful in 
reproducing the observed bidirectional motion of Dictyostelium endosomes. 
 
We performed Monte Carlo simulations of bidirectional transport using 
the above model as follows. The cargo with a certain
fixed number of attached motors (in the range $D=1-15$ and $K=1-3$)
started motion in a certain initial state,  which was usually
$|10\rangle$. Individual motors are free to attach to and detach from
the filament with the rates as mentioned above, thus facilitating
changes in the state of the cargo. The cargo moves in the plus
direction with a speed $v_+$ in $|0k\rangle$ states, where $k>0$, and
in the minus direction with speed $v_{-}$ in the $|d0\rangle$ states,
where $d>0$.  All $|dk\rangle$ states with $d>0,k>0$ represent
TOW situations in which the cargo velocity is small in comparison $v_{\pm}$ (Some typical values are $v_c(1,1)=0.235 \mu$ms$^{-1}$, $v_c(2,1)$=97.07nm s$^{-1}$, $v_c(3,1)$=44.65nm s$^{-1}$ etc.). Such a TOW situation ends when 
one set of motors  give up under the opposing
load from the other set. The cargo detaches completely from the
filament the moment it assumes the $|00\rangle$ state.

A few simulations were also performed with only similar motors (eg. $K=0$), so 
as to reproduce the stalling of a bead coated with a definite number $n$ of dyneins 
(or kinesins) in an optical trap. In this case, we assume that an opposing load 
$F$ acts on the bead, whose effect is reflected in the enhanced detachment rate of the motors. 
The next section discusses how the various single molecule
parameters to be used in the simulations were extracted from {\it
  vitro} experimental data reported in \cite{SOPPINA}.

\subsection{Analysis of {\it in vitro} data and extraction of
  single molecule parameters}

In this section, we describe the procedure by which single molecule
parameters such as the intrinsic dissociation rates $\epsilon^{0}_{\pm}$,
binding rates $\pi_{\pm}$ and detachment forces $f_{\pm}^{d}$ (listed
in Table \ref{tab:tab3}) were extracted from the analysis
of {\it in vitro} experimental data reported in Soppina
et. al.\cite{SOPPINA}.

It is useful here to introduce the notion of mean first passage time
(FPT) $T_{\pm}^{n}(F)$ of an assembly of similar motors, attached to a cargo,
which is simply the mean time taken for such a cargo to completely detach
from the filament, assuming that the starting state is $|{n}\rangle$,
and an external opposing load $F$ acts on the cargo throughout this
time. In optical trap experiments, $F$ is the force exerted by the 
trap on the bead, while in in vivo experiments, this force is produced by opposing motors. 
The general expression for this quantity was derived by Klumpp
and Lipowsky\cite{KLUMPP} (and also in \cite{MJMULLER}), which we use in the following analysis. 

Let us consider the no-load situations first, and put $F=0$. In this
case, the FPT is most easily obtained from the mean run-lengths
$\ell_{\pm}^n$ measured experimentally. Since cargo velocities
$v_{\pm}$ do not significantly depend on motor numbers, these
  quantities are related as

\begin{equation}
T_{\pm}^n(0)=\frac{\ell_{\pm}^n}{v_{\pm}} 
\label{eq:eq3}
\end{equation}

For one motor, $T_{\pm}^1(0)=1/\epsilon_{\pm}^{0}$, therefore, the
measured run-lengths immediately give the intrinsic single molecule
detachment rates under no load. We now consider a cargo moved by a two-motor assembly. The FPT in this case has the general form (as derived in
\cite{KLUMPP})

\begin{equation}
T_{\pm}^2(F)=\frac{1}{\epsilon_{\pm}(F)}+\frac{1}{\epsilon_{\pm}\left(F/2\right)}+\frac{\pi_{\pm}-\epsilon_{\pm}(F)}{2{\epsilon_{\pm}\left(F\right)}~{\epsilon_{\pm}\left(F/2\right)}}
\label{eq:eq4}
\end{equation} 

Putting $F=0$, we may therefore use the FPT above to deduce the binding
rates $\pi_{\pm}$, after using the experimentally measured two-motor run
lengths in Eq.\ref{eq:eq3} to obtain $T_{\pm}^2(0)$.

Let us now consider the case $F>0$. In \cite{SOPPINA}, the stall time
for beads coated with 1,2 and 3 dyneins (and also 1 and 2 kinesins) 
were measured directly (reproduced in Table 1). For $n=1$, the stall time therefore directly gives
$\epsilon_{\pm}(F_{\pm}^s)$, where $F_{\pm}^s$ is the stall force for
each species. By directly substituting into Eq.\ref{eq:eq1} and Eq.\ref{eq:eq2}, we may
obtain the detachment forces $f_{\pm}^d$. 

The results of the analysis of this section are summarized in Table \ref{tab:tab3}. 

\begin{table}
\begin{tabular}{|c|c|c|c|}
\hline Molecular motor& Intrinsic unbinding rate
($\epsilon_{\pm}^0$)& Binding rate  ($\pi_{\pm}$) & Detachment
force   $f_{\pm}^d$\\
\hline &&& \\ Dynein (-)&0.667 $s^{-1}$&2.740 $s^{-1}$&  0.546pN\\\hline &&&
           \\Kinesin(+)&0.314 $s^{-1}$& 0.904 $s^{-1}$&  5.169pN\\ \hline
\end{tabular}
\caption{List of single molecule parameters extracted from {\it in
    vitro} experiments}
\label{tab:tab3}
\end{table}

\subsection{Load-dependent detachment rate}

In the course of a TOW between a single, strong kinesin and several weak
dyneins, it is quite likely that there will be many intermediate
dynein detachment events. When less than 5 dyneins hold onto the
filament, against an opposing load of a kinesin, the individual
dyneins (under equal load-sharing assumption) are forced to a
situation with super-stall loads. Therefore, a quantitative
understanding of the TOW events requires knowledge of the
detachment rate of dynein under super-stall load conditions. Although direct 
experimental evidence is available now\cite{KUNWAR1,MCKINNEY}. it is useful to obtain 
insights into this issue by considering the stall time of
multiple motor coated beads under a load equaling the combined
stall force of all the motors\cite{SOPPINA}. For the two-motor case, the
experimentally determined stall time is 0.4 seconds. However, when the
FPT for this system is computed from Eq.\ref{eq:eq4} using the
exponential detachment rate in  Eq.\ref{eq:eq2}, the result turns out to
be $\sim$ 0.133 seconds, which is even less than the one-motor stall
time (0.2s). 

%Unlike
%kinesin\cite{SOPPINA,COPPIN}, only one experiment appears to
%have probed the dissociation of dynein in the super-stall
%regime\cite{MCKINNEY} (see more below). 

The reason for the above discrepancy is not hard to
understand. Let us first assume that the binding rate of dynein is
small (in comparison with its detachment rate). Given that a single dynein will detach within 0.2 s under its
stall load, one dynein in a team of two will detach (on an
average) in half as much time, i.e, 0.1 s (since the effective
detachment rate doubles), which puts the still attached 
dynein under a load equal to twice its stall force. Therefore, if we
apply the rate in Eq.\ref{eq:eq2}, the 
second dynein will detach within a time much less than 0.2 s, 
consistent with the result obtained here. In order for this not to
happen, the binding rate will have to be sufficiently high so that the
detached first dynein will manage to rebind before the second dynein
also detaches. However, the binding rate thus obtained from Eq.\ref{eq:eq4} turned out to be too high (102.52$s^{-1}$), 
in comparison with the value estimated in the previous section from run-length 
measurements. 

A second, and related possibility is that there may be other, non-filament-bound motors coated on the bead at
any given point of time. In this case, when one of the bound motors unbinds, the bead may possibly tilt from its original position 
(relative to the filament), allowing one of the other motors to bind\cite{YU}. In this case, therefore, we are dealing with a situation where we have a total of 
$n$ motors on the bead, but only two can possibly bind to the filament at any point of time (if this were not the case, we would have observed 
a stalled two-motor bead moving away from the trap center sometimes, which was never seen). However, in order to reproduce the observed 
stall times, again, the individual binding rate of each motor would have to be too high (eg. $\pi_{-}\simeq 50$ s$^{-1}$, 33 s$^{-1}$ and 
25 s$^{-1}$ for $n=3,4$ and 5 respectively).

We, therefore, considered the possibility that the second dynein,
under super-stall load, may not detach as fast as predicted by
Eq.\ref{eq:eq2}. As the simplest conjecture, we recalculated the FPT
using a detachment rate that increases exponentially with load up to
the stall force, but saturates at the stall-force level for
super-stall loads (motivated partly by experimental data for kinesin \cite{COPPIN}), i.e., 

\begin{eqnarray}
f_{-}(\lambda)=\exp\left(\frac{\lambda}{f^d_{-}}\right)~~~~\lambda\leq F_{-}^s\nonumber\\
f_{-}(\lambda)=\exp\left(\frac{F_{-}^s}{f^d_{-}}\right)~~~~\lambda >F_{-}^s
\label{eq:eq5}
\end{eqnarray}

With this form for the detachment rate of dynein, and the binding rate
in Table 1, the FPT for the two-dynein coated plastic bead under a load $F=2F_{-}^s$
turned out to be $\sim$ 0.353s, which is much closer to the experimental value. 

Only a few experiments have directly measured detachment rate of dynein at super-stall loads. In \cite{MCKINNEY}, it was 
reported that a single dynein survives only about 0.94 s under a load of 2pN. This does not agree with 
Eq.\ref{eq:eq5} above (which predicts 0.2s for survival time), while it seems consistent with Eq.\ref{eq:eq2} which predicts 38ms. 
More recent experiments reported in \cite{KUNWAR1} shows a marked decrease of detachment rate with load in the super-stall regime, in contrast to 
the sub-stall regime where it increases with load. The detachment rate in Eq.\ref{eq:eq5} is broadly consistent with the data in \cite{KUNWAR1}, except 
for the peak near the stall force.  It is also important to note that the abrupt
change in the detachment rate at stall load as suggested by
Eq.\ref{eq:eq5} need not be taken literally; a smoother change between
sub and super-stall loads (eg., tanh function replacing exponential in Eq.\ref{eq:eq2}) also produces similar results.

%The following comments are in order here. In recent optical trap
%experiments, it was observed that a a single dynein survives only about
%94ms under a load of 2pN\cite{MCKENNEY}. This does not agree with
%Eq.\ref{eq:eq5} above (which predicts 0.2s for survival time), while
%it seems consistent with Eq.\ref{eq:eq2} which predicts 38ms.
%Therefore, it could be that the modified detachment rate suggested in Eq.\ref{eq:eq5} is an {\it
%  effective single molecule property}, which applies 
%only when dyneins act in a team, and not for single, isolated
%motors. It is also important to note that the abrupt
%change in the detachment rate at stall load as suggested by
%Eq.\ref{eq:eq5} need not be taken literally; a smoother change between
%sub and super-stall loads (eg., tanh function replacing exponential in
%Eq.\ref{eq:eq2}) also produces similar results.

\begin{figure}
\centering
\vspace{0.5cm}
\includegraphics[angle=-90,width=0.5\linewidth]{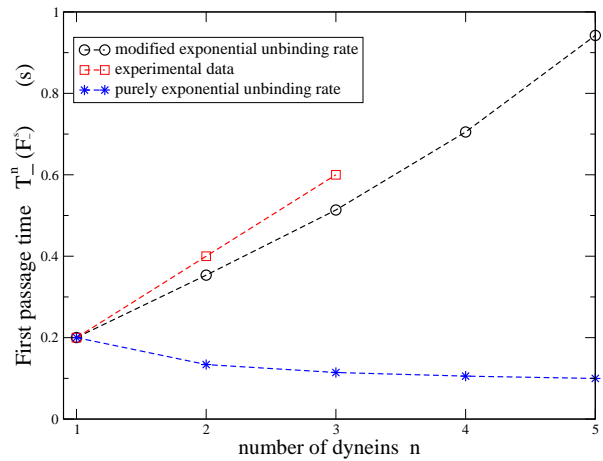}
 % fpt.eps: 612x792 pixel, 72dpi, 21.59x27.94 cm, bb=0 0 612 792
\caption{Experimental stall time data(squares, red) of dyneins
  \cite{SOPPINA} and FPT with
saturating exponential (ref. Eq.\ref{eq:eq5})unbinding rate(black
 circle) are plotted as a function of number of dyneins. FPT with purely 
exponential (Eq.\ref{eq:eq2}) load dependence(stars, blue) shows clear mismatch with the the 
experimental data.} 
\label{fig:fig1}
\end{figure}

To compute the FPT for beads with 3 and 4 motors attached, we resorted
to numerical simulations. The FPT was computed as explained in the
last section, and the results are plotted in Fig.\ref{fig:fig1}. For
comparison, we have also shown the corresponding results with the
exponential load-dependent detachment rate. It is clear that the {\it
  saturating exponential} form of detachment rate in Eq.\ref{eq:eq5}
agrees much better with observations, in comparison with the purely
exponential form in Eq.\ref{eq:eq2}.

What about the detachment of kinesins at super-stall loads? It is reported in \cite{SOPPINA}
that a single kinesin can sustain a super-stall
load of  7.8 pN for about 0.7 seconds. Using the intrinsic detachment
rate, binding rate and detachment force for kinesin extracted from
experiments (Table 3), the exponential rate in Eq.\ref{eq:eq2} gives a
detachment rate of almost exactly 0.7 s, which indeed matches with the
experimental value! It appears likely, therefore, that for kinesin,
the form in Eq.\ref{eq:eq2} may be appropriate at least for loads less
than 7.8pN. Although the behaviour at much larger loads is unknown, and since 5-6 dyneins are enough to stall a kinesin 
in TOW, we assumed an exponential detachment rate for kinesin, i.e., 

\begin{equation}
f_{+}(\lambda)=\exp\left(\frac{\lambda}{f^d_{+}}\right).
\label{eq:eq5+}
\end{equation}

Eq.\ref{eq:eq5} and Eq.\ref{eq:eq5+} will be used in our simulations
described in the next section.

\subsection{Numerical simulations of bidirectional transport of endosomes}

Having established the mathematical forms for the functions $f_{\pm}(\lambda)$ to agree 
with {\it in vitro} stall time data, we now proceed to model the situation where a set of kinesins and dyneins 
pull a cargo in opposite directions. 

\begin{figure}
\centering
\includegraphics[angle=-90,width=0.5\linewidth]{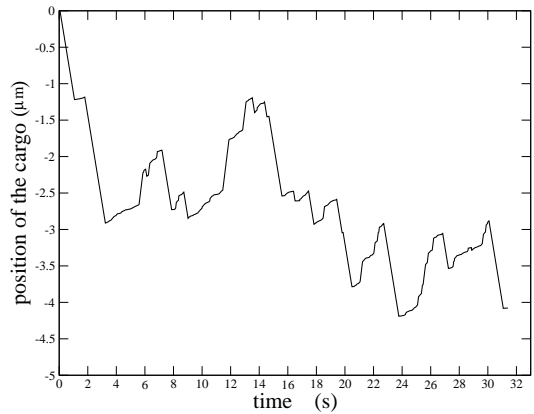}
\caption{A representative trajectory of cargo pulled by 1 kinesin against 5
   dyneins. Trajectory shows frequent TOW phases, and long
   negative runs, similar to that reported in \cite{SOPPINA}.
}
\vspace{1.0cm}
\label{fig:fig2}
\end{figure}
 
We now used the parameters listed in Table \ref{tab:tab3} and the detachment rates in 
Eq.\ref{eq:eq5} and Eq.\ref{eq:eq5+} to simulate the motion of an
endosome. A single run was continued until the cargo detached from the
filament, and various characteristics of the motion (duration of
plus/minus runs, duration of TOW events, statistics of reversals etc.)
were recorded. For each set of values ($D,K)$, a total of $5\times
10^5$ independent runs of the cargo were used to compute averages. 
 
\begin{figure}
\vspace{1.0cm}
\centering
\includegraphics[angle=-90,width=0.5\linewidth]{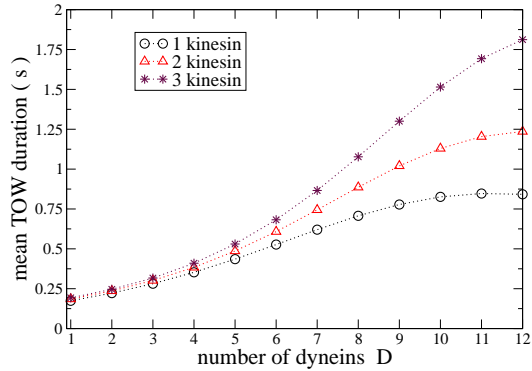}
\caption{The mean TOW duration increases as we increase dynein
  numbers, until the point where the kinesin(s) detach under the load
  of opposing dyneins. In this curve, we have used Eq.\ref{eq:eq5} for the detachment rate of
  dynein, while the exponential form in Eq.\ref{eq:eq5+} is used for
  kinesin.}
\vspace{1.0cm}
\label{fig:fig3}
\end{figure}

A typical trajectory of a cargo being pulled by five dyneins against one
kinesin is shown in Fig.\ref{fig:fig2}. As observed in experiments, frequent pauses
in the course of motion are seen, indicating TOW situations. The
mean duration of the TOW was found to be in the range 0.43, 0.52 and
0.62 second for one kinesin against 5, 6 and 7 dyneins
respectively. These numbers are somewhat smaller, but comparable to
the experimental value of $\sim 1$ second (Table \ref{tab:tab2}). TOW
durations for cargo dragged by 
2 kinesins against 10 dyneins and 3 kinesins against 9 and 10  dyneins are similar to what
was reported recently in human adenovirus transport(2-3 sec), and the
total number of motors involved also agrees with the optimal number
estimated in these experiments ($\sim$14) \cite{ADENO}. After an initial gradual
increase with increasing dynein number, the TOW duration(for fixed
number of kinesins) appears to saturate to a value which depends on
the number of opposing kinesins, as depicted  in
Fig.\ref{fig:fig3}. 

The trajectories observed in simulations also show long negative runs and short positive runs as in the experiments.
However, the time-scales here are vastly different: whereas the
experimentally observed mean duration of uninterrupted minus-directed and plus-directed runs
are nearly 10 s and 5 s respectively (Table \ref{tab:tab2}), the corresponding numbers in the
simulations were 1.099s and 0.075s (i.e., for 1 kinesin against 5
dyneins, see Table \ref{tab:tab4}). However, in the experiments, only
those 
TOW events were reported which produced a
change of direction of motion of the cargo; short-lived TOW events which failed to
produce a reversal in direction were not reported. In order to
accommodate this possibility, we carried out a coarse-graining of
the trajectories where all pause events which did not result in a
change of direction were not counted, i.e., these events were simply
absorbed into the runs, which are now called {\it trips}(\cite{WELTE}), to
distinguish them from the runs. With this procedure, the mean minus trip time
changed very little from the corresponding run time, eg., from 1.099 s to
1.302s for 5 dyneins, but the plus trips were significantly longer than
the runs, with the mean time increasing almost four-fold, to
0.307s (Table \ref{tab:tab5}). However, these numbers still significantly fall short of the
experimental values. This issue is discussed in more detail in the
last section. 

\begin{figure}
\vspace{1.0cm}
\centering
\mbox{\subfigure[]{\includegraphics[angle=-90,width=0.4\linewidth]{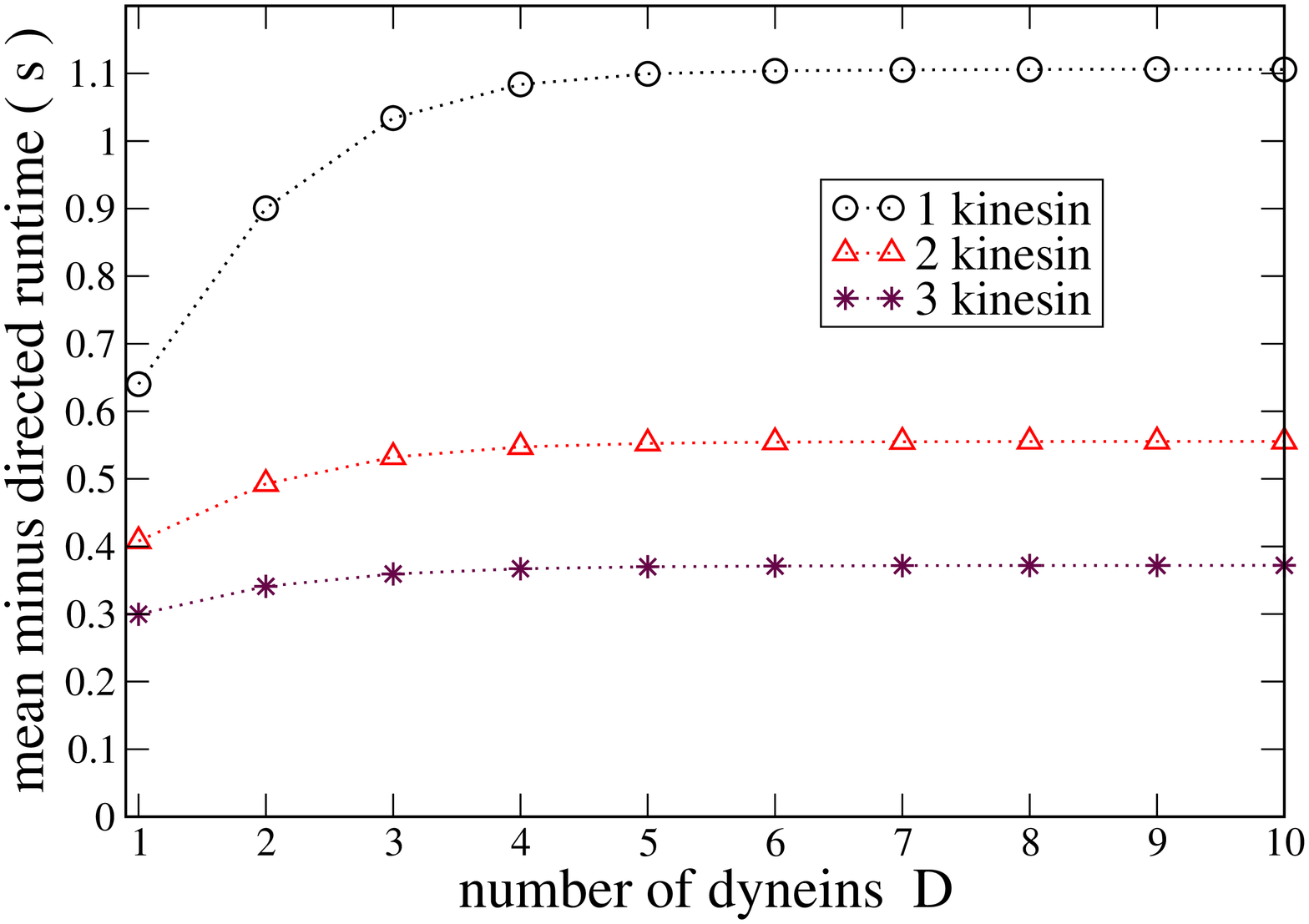}}\quad
\subfigure[]{\includegraphics[angle=-90,width=0.4\linewidth]{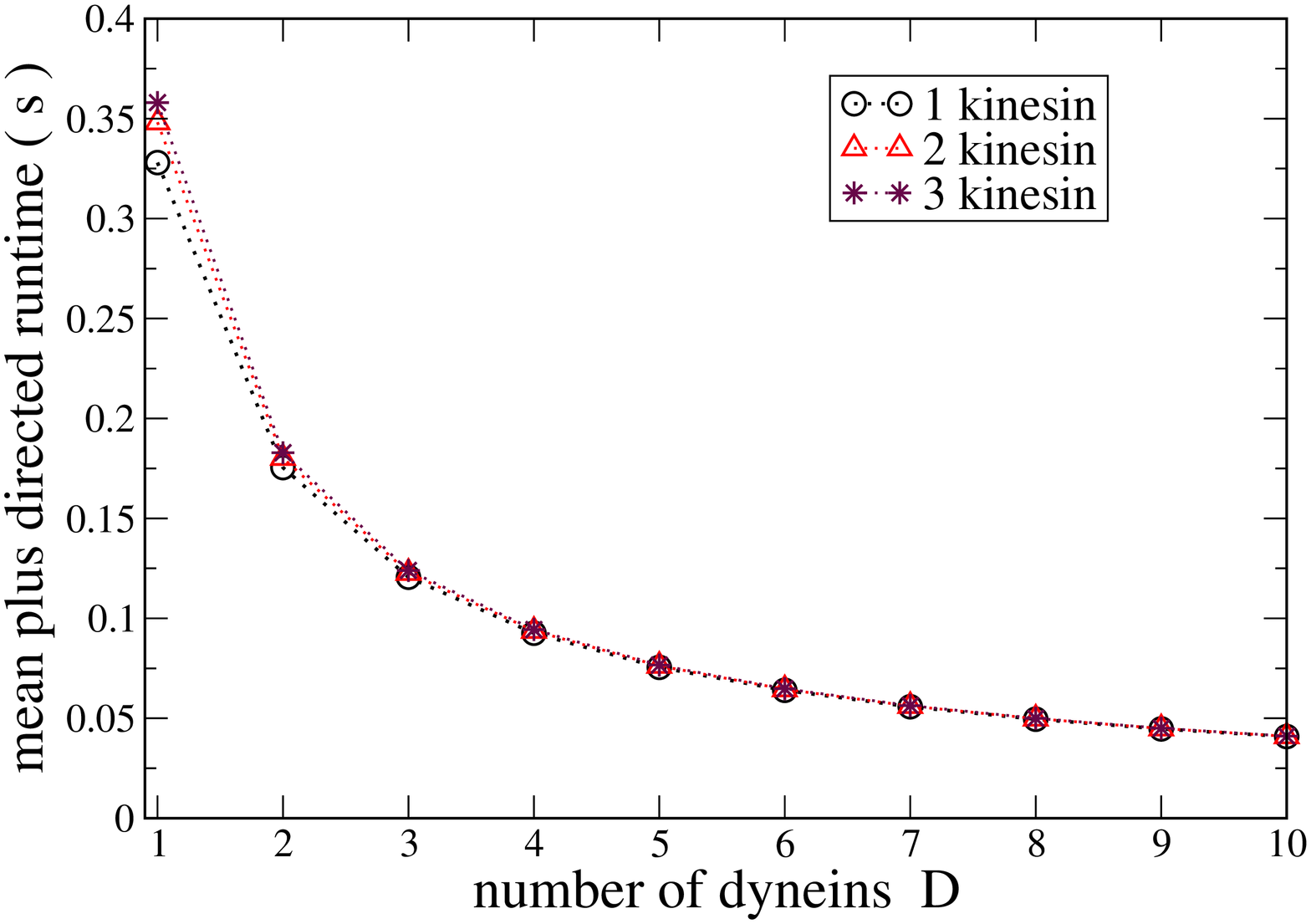}
}}
\caption{The mean run-length in the minus(a) and plus(b) direction for
varying numbers of kinesin and dynein.}
\label{fig:fig4}
\end{figure}

\begin{table}
\begin{tabular}{|c|c|c|c|c|c|}
\hline
Kinesin&Dynein& distance covered ($\mu m) $&TOW(s)&minus 
run(s)&plus run(s)\\ \hline 
  &     4&    -6.84&     0.35&    1.083&    0.092\\
  &     5&   -14.36&     0.43&    1.099&    0.075\\
 1&     6&   -26.44&     0.52&    1.103&    0.064\\ 
  &     7&   -46.08&     0.62&    1.105&    0.055\\\hline

  &     7&   -12.10&     0.74&    0.554&    0.056\\
  &     8&   -25.26&     0.88&    0.555&    0.049\\
 2&     9&   -47.06&     1.02&    0.555&    0.044\\
  &     10&  -83.83&     1.12&    0.555&    0.040\\\hline
\end{tabular}
\caption{A summary of results from numerical simulations of endosome
  transport. The load-dependent detachment rates are taken from
  Eq.\ref{eq:eq5} and Eq.\ref{eq:eq5+}}
\label{tab:tab4}
\end{table}

\begin{table}
\begin{tabular}{|c|c|c|c|c|c|}\hline
Kinesin&Dynein& distance ($\mu m)$&TOW (s)& minus trip (s) &plus trip (s)\\ \hline
  &    5&    -14.36&    0.51&	    1.302&    0.307\\
 1&    6&    -26.44&    0.61&	    1.393&    0.304\\
  &    7&    -46.08&    0.71&	    1.493&    0.293\\\hline

  &    7&    -12.10&    0.89&	    0.908&    0.467\\
  &    8&    -25.26&    1.06&	    1.039&    0.471\\
 2&    9&    -47.06&    1.22&	    1.186&    0.452\\
  &   10&    -83.83&    1.37&	    1.332&    0.411\\\hline
\end{tabular}
\caption{The table shows `coarse-grained' results of numerical
  simulations, where TOW situations without direction reversals
  are absorbed into the preceding run. In this table, TOW only refers
  to those events which are accompanied by direction reversals, and
  runs are extended to trips (see \cite{WELTE}).}
\label{tab:tab5}
\end{table}

\begin{figure}
\vspace{0.5cm}
\centering
\mbox{\subfigure[]{\includegraphics[angle=-90,width=0.4\linewidth]{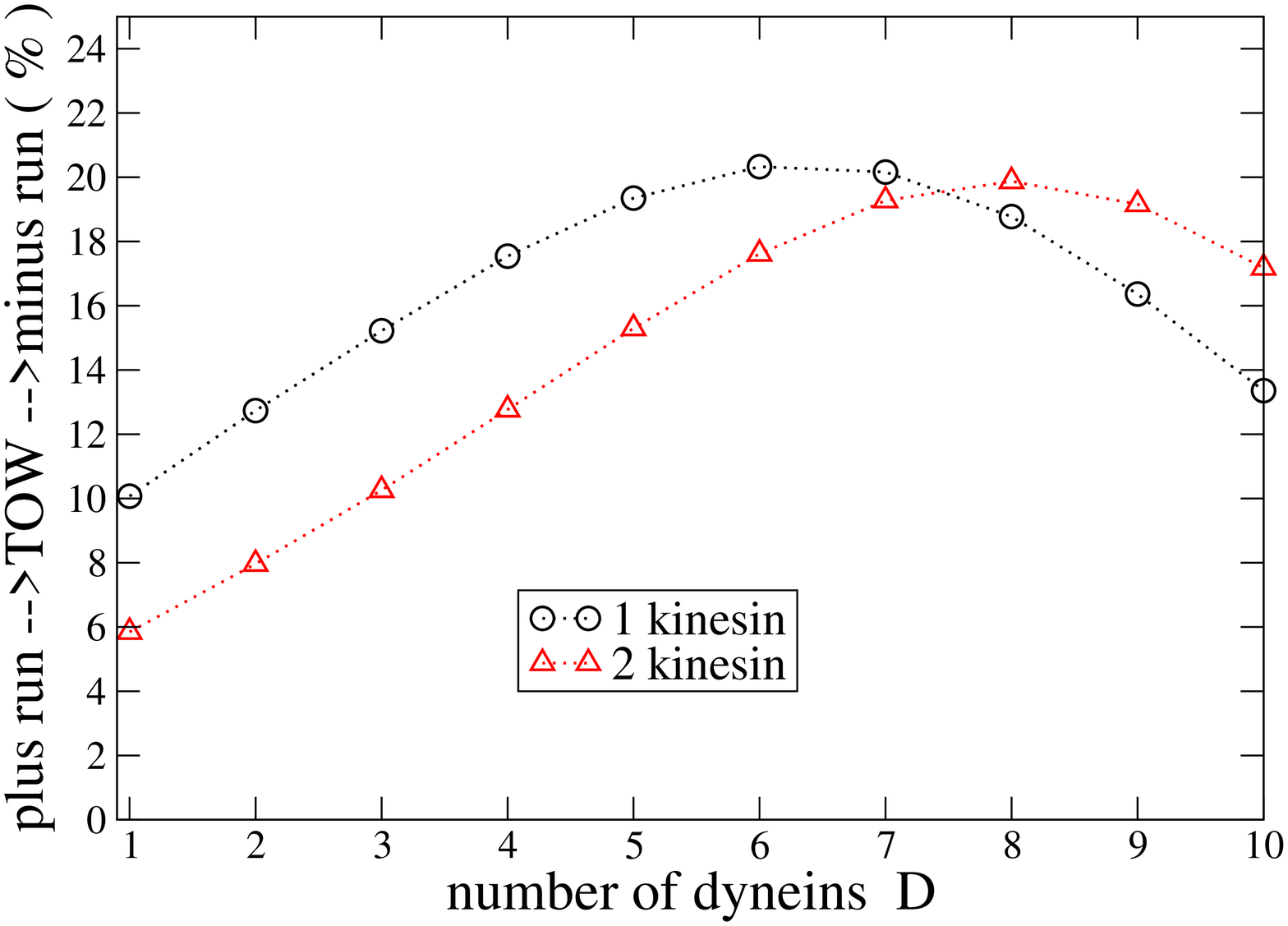}}\quad
\subfigure[]{\includegraphics[angle=-90,width=0.4\linewidth]{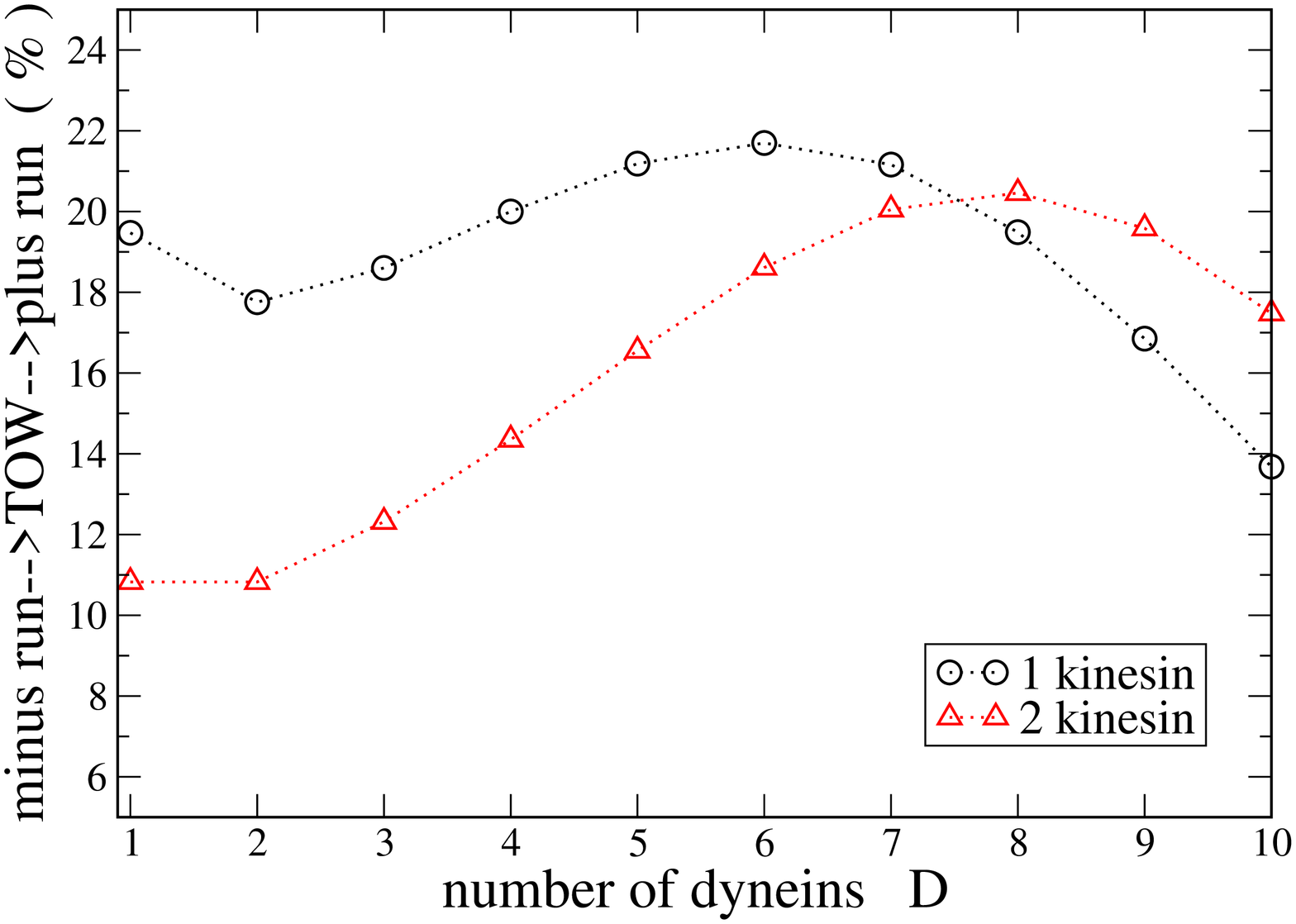}
}}\vspace{1.0cm}

\mbox{\subfigure[]{\includegraphics[angle=-90,width=0.4\linewidth]{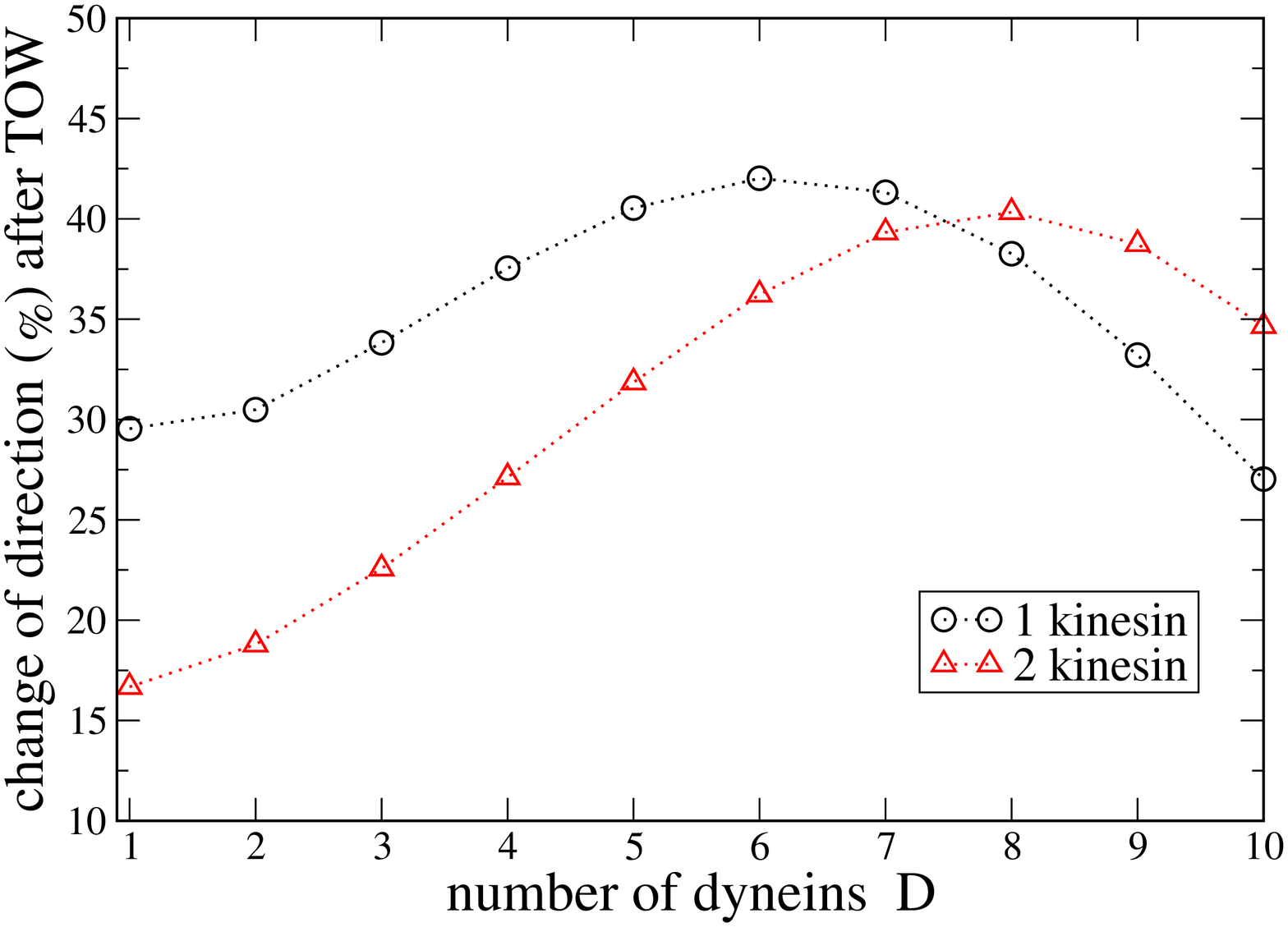}}\quad
}
\caption{The figure shows the fraction of TOW events that resulted in
  a reversal in direction, when the TOW was preceded by (a)plus run
  (b) minus run and (c) either. The data in (c) is therefore simply
  the sum of that in (a) and (b).  
}
\label{fig:fig5}
\end{figure}

{\it Direction reversals}: From the simulations, we determined the fraction of 
TOW/pause events that culminates in a change of direction, separately for TOW 
events that were preceded by a minus directed run and a plus directed run. 
In Fig.\ref{fig:fig5}, this fraction is plotted as a function of the number 
of dyneins, for a fixed number of opposing kinesins ($K=1$ or 2), for TOWs 
following a plus run (Fig.\ref{fig:fig5}a), minus run
(Fig.\ref{fig:fig5}b) and the sum(Fig.\ref{fig:fig5}c) of the two
separately. Interestingly, we observe that a cargo moving in the plus
direction is more likely to continue moving in the same direction
after a TOW event, and the same holds true for minus directed runs
also. Therefore, the bidirectional cargo motion as a whole, with
interspersed plus and minus run segments, may be 
viewed as a {\it persistent random walk}.

For $K=1$, a maximum fraction of reversals (42.02\%, see Fig.\ref{fig:fig5}c) irrespective of
the direction of the initial runs, is observed at $D=6$ whereas for $K=2$, the maximum shifts 
to $D=8$. The maximization of reversals is likely to be a strategy to 
effectively ``slow down'' the cargo so as to increase the number of TOW events 
before detachment, and therefore the probability of fission. It is interesting 
that the optimal combination (6 dyneins for 1 kinesin) is close to the experimentally 
determined ratio from optical trap studies\cite{SOPPINA} (5-6 dyneins against a kinesin). 

\begin{figure}
\centering
\includegraphics[angle=-90,width=0.5\linewidth]{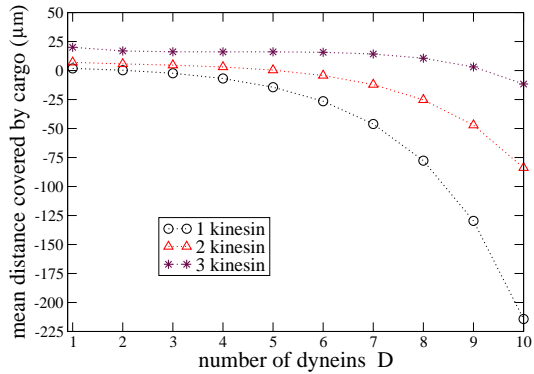}
\caption{The figure shows the mean total distance moved by the cargo
  before its eventual detachment from the filament. With the present
  parameters, 5-6 dyneins are required (against 1 kinesin) to move the
  cargo over a distance of $\sim 20\mu$m in the minus direction. 
}
\vspace{1.0cm}
\label{fig:fig6}
\end{figure}

{\it Total distance covered before detachment}: We also determined the
mean total distance traveled by a cargo before its eventual detachment from
the filament (re-bindings of the cargo to the filament were not considered). In Fig.\ref{fig:fig6}, we plot the
mean displacement of the endosome from its initial position as a
function of the number of dyneins $D$, for fixed number of kinesins
$K$.  For $K=1$, the mean displacement is positive for $D=1$ and 2,
whereas it is always in the minus
direction for $D>2$. The displacement becomes appreciable (more than 10$\mu$m) only for $D>4$. For large $D$, the net
displacement appears to increase exponentially with $D$.

\begin{figure}
\vspace{0.5cm}
\centering
\mbox{\subfigure[]{\includegraphics[angle=-90,width=0.4\linewidth]{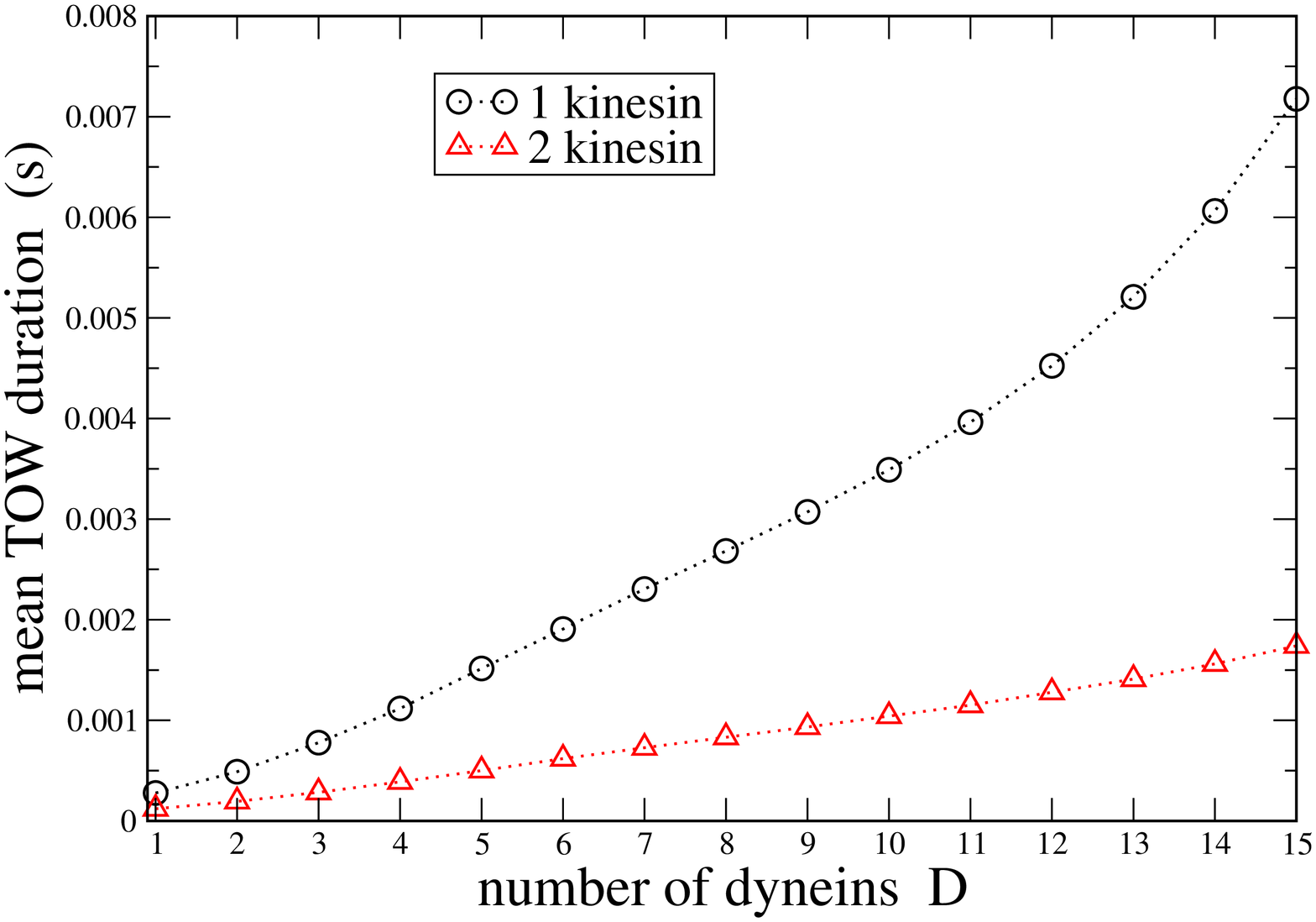}}\quad
\subfigure[]{\includegraphics[angle=-90,width=0.4\linewidth]{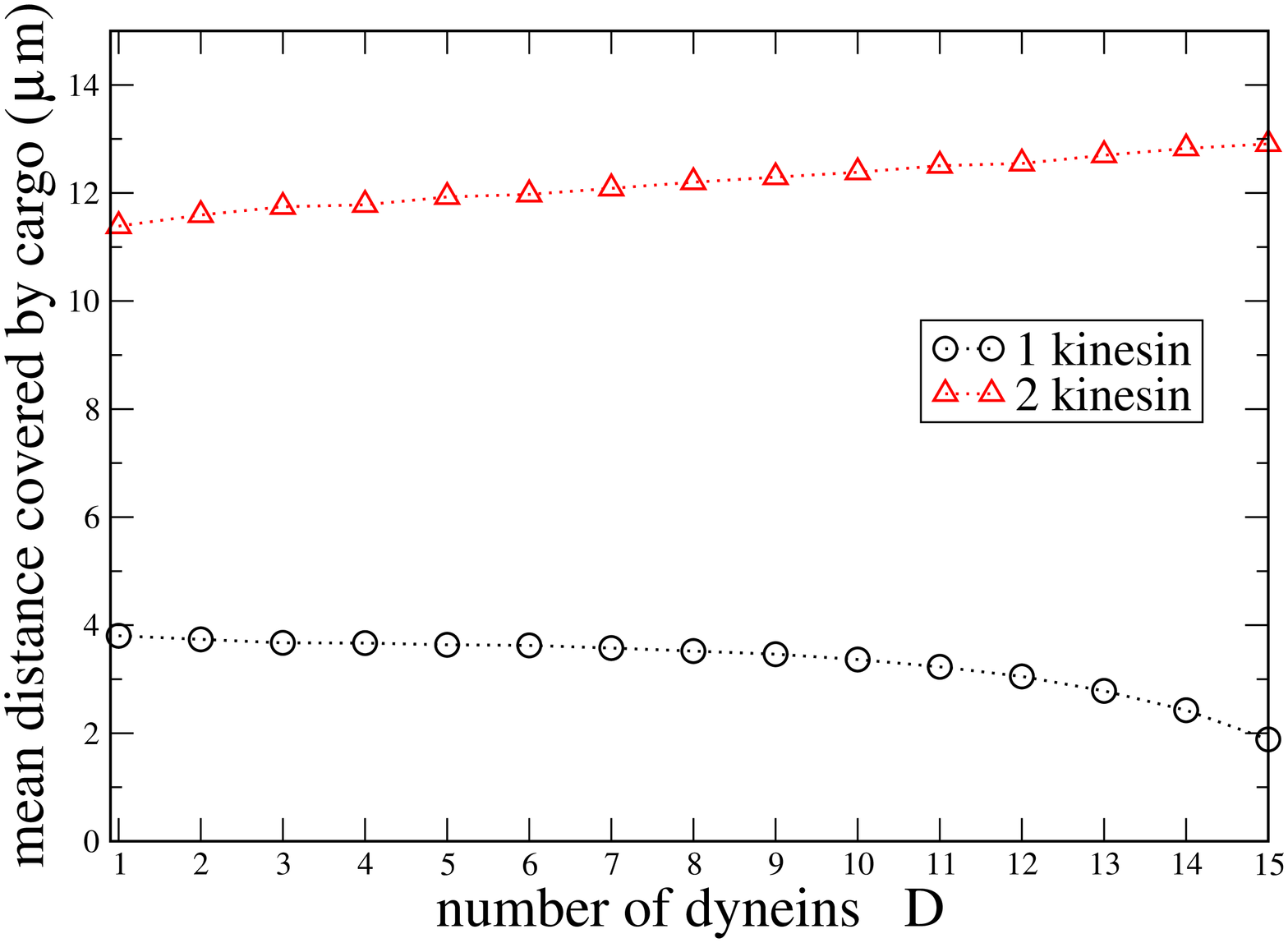}
}}

\caption{
Purely exponential load dependence (used for both dynein and kinesin)
of unbinding rate shows mismatch with the experimental
results. (a) TOW duration  is much less than what is observed in
experiments (Table \ref{tab:tab2}). (b)The mean net displacement of
the cargo is generally positive here, while endosomes
in the cell are biased to move in the minus direction on an average.
}
\label{fig:fig7}
\end{figure}

{\it What if dynein detachment rate was exponential for all loads?} 
Fig.\ref{fig:fig7} shows the mean TOW
duration in this scenario, for varying number of motors. The TOW
lasts typically only less than a hundredth of a second in this case!
Furthermore, the net transport is seen to be in the plus
direction for up to 15 dyneins against a kinesin. Neither of these
results agree with experimental observations, which effectively rules
out the purely exponential unbinding rate. 

\begin{figure}
\vspace{0.5cm}
\centering
\includegraphics[angle=-90,width=0.5\linewidth]{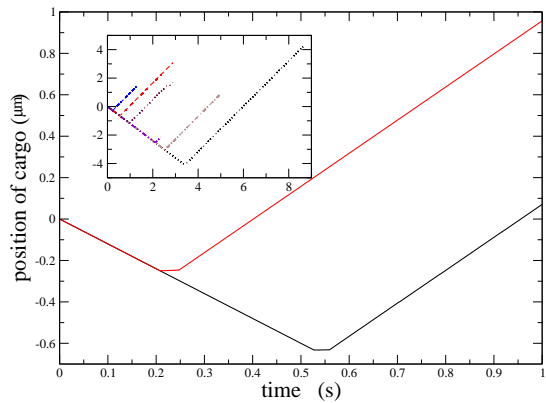}

\caption{Trajectories of endosomes driven by one kinesin and five
  dyneins, using the purely exponential form of the detachment rate in
  Eq.\ref{eq:eq2} for both dyneins and kinesin. Plus run to minus
  direction reversals can be seen rarely(lass than 0.03\%) while minus
to plus reversals(2.27\%) are much less than what is seen in Fig.\ref{fig:fig5}.}\hspace{1cm}
\label{fig:fig8}
\end{figure}

\section{Discussion}

Multiple-motor driven bidirectional transport of cellular cargoes is now established as a
widespread phenomenon across a variety of cells. In this paper, we
have carried out a numerical simulation study of a model of endosome transport in
{\it Dictyostelium} amoeba reported in \cite{SOPPINA}, in which oppositely pulling dyneins and
kinesins engage in a TOW which results in bidirectional
transport and occasional fission of the endosome. Our objective was to see how well the
observed kinetic features of the bidirectional transport are
consistent with (a) single-molecule parameters (binding/unbinding
rates) that characterize these motors measured {\it in vitro} (b)
measurements of stall-time of individual motors and motor teams in
optical traps, and (c) the commonly assumed exponential dependence of
detachment rate of a motor on the opposing load. 

Our analysis, supplemented with explicit numerical simulations show
that the mean duration of the TOW events is consistent with
experimental data, provided we assume a somewhat modified form for the
load-dependence of the effective single molecule detachment rate of dynein. Essentially, this
requires that the detachment rate saturates at super-stall loads,
while it increases with load at sub-stall loads. The loss of
sensitivity at super-stall loads helps the dynein team to hold their
ground against a stronger kinesin in a TOW situation, where the
competition is frequently unfair as far as balance of forces is
concerned. The dependence of detachment rate on load suggested in this paper is similar to 
what was found in recent direct optical trap measurements\cite{KUNWAR1}.

\begin{table}
\begin{tabular}{|p{1.5cm}|p{1.7cm}|p{1.7cm}|p{1.7cm}|p{1.7cm}|p{1.7cm}|p{1.7cm}|}\hline
 $\pi_{-}(s^{-1})$&\multicolumn{2}{c|}{TOW duration $(s)$}&\multicolumn{2}{c|}{$T_{-} ~~~(s) $}&\multicolumn{2}{c|}{$T_{+} ~~~(s) $}\\\cline{2-7}
      $ $         & before coarse-graining&after coarse-graining &run\hspace{1cm} duration&trip\hspace{1cm} duration&run\hspace{1cm} duration&trip\hspace{1cm} duration\\ \hline   
0.005&	0.175&	0.175&	1.069&	1.076&	  2.931&	2.941\\	
0.2&	0.198&	0.217&	1.637&	1.653&	  0.764&	0.884\\	 
0.4&	0.216&	0.250&	2.134&	2.161&	  0.435&	0.578 \\
0.6&	0.233&	 0.279&	2.524&	2.563&	  0.306&	0.462 \\
0.8&	0.251&	0.305&	2.810&	 2.863&	  0.236&	0.403 \\
1.0&	0.269&	0.330&	3.012&	3.080&	  0.192&	0.368 \\
2.0&	0.364&	 0.444&	3.400&	3.544&	  0.101&	0.314\\
3.0&	0.466&	0.550&	3.476&	3.704&	  0.069&	0.306\\
\hline\hline
\end{tabular}\caption{The table shows results of simulations with a
  reduced value of $\pi_{+}=0.285s^{-1}$, corresponding to a two-motor
  run length of $l_{+}=10\mu m$, and $\pi_{-}$ varied in the range
  shown. The results for runs and trips are shown
  separately. Interestingly, increasing $\pi_{-}$ increases $T_{-}$,
  but at the same time, reducing $T_{+}$, so an optimal value could
  not be identified.}
\label{tab:tab6}
\end{table}

The principal weakness of the present model is its apparent failure to reproduce simultaneously all the observed features of endosome transport in a consistent manner. In particular, we see a clear mismatch between the binding rates deduced from observed in vitro run lengths (with similar motors) and the termination rates of unidirectional runs between TOW events during bidirectional motion.  Some of the possible reasons for this discrepancy are discussed below, within the framework of our model.

\begin{enumerate}

\item It is likely that crowding of motors on the endosome prevents transiently unbound motors from
rebinding to the filament at their natural rate, so that
unidirectional runs continue longer than they normally should. Indeed,
a recent computational study\cite{YU} lends some support to this argument. 
In order to verify how robust the tug of war duration is, against
changes in the binding rates of the motors, we 
carried out simulations with reduced values of both $\pi_{+}$ and
$\pi_{-}$, while keeping the intrinsic dissociation rates the same. 
Reducing $\pi_{+}$ to 0.285s$^{-1}$
significantly increased the minus run duration without affecting TOW;
however, a similar reduction in $\pi_{-}$ reduced the TOW duration
{\it and} the minus run intervals, while extending the plus runs (see
Table \ref{tab:tab6}). While the effect of $\pi_{-}$ on TOW and plus
runs is obvious, the effect on minus run duration is a non-trivial
feature. No single set of values of ($\pi_{+},\pi_{-}$) was found to reproduce the experimental
numbers for all the three important quantities, i.e., plus run/trip,
minus run/trip and TOW duration (data not shown), so this could well be a limitation of
the model itself. 

\item There is a likelihood that motors may detach from and re-attach to the
cargo many times within the time scale of observation of transport. In this scenario,
during a minus run, frequently, a kinesin  may not be present on the
cargo at all, and the run is interrupted only when a kinesin first
binds to the cargo from solution which then binds to the
filament. Therefore, the duration of the plus(minus) run is determined
by the (presumably smaller) rate of binding of the dynein(kinesin) to the cargo from
cytoplasm. In principle, therefore, this could result in a minus run being extended well beyond the time-scale determined 
by the binding rate of kinesin. However, it is unclear how this will affect plus runs, as the number of dyneins on the cargo 
is typically much larger than one.

\item Finally, the possibility that a coordinating complex may indeed be responsible for regulating the
duration of the runs while active TOW dictates reversals cannot be completely ruled out either.

\end{enumerate}

Among the three possibilities discussed above, we have ruled out (i) after extensive simulations, while a systematic exploration of (ii) will require significant extension of the present model with additional data, well beyond the scope of the present paper. In our opinion, (iii) should be considered seriously after exhausting the other options. At this time, we feel that more experiments on bidirectional transport with detailed measurements of run lengths, pause/TOW events and reversal statistics would be desirable for a better quantitative understanding and characterization of this intriguing phenomenon. In addition, an extension of the present mean-field model to include the motor-motor and motor-cargo interactions\cite{KUNWAR,NEW2,BOUZAT1,BOUZAT2} could also provide additional insights.

%From the modeling perspective, the principal limitation of the
%present `mean-field' model is the presupposition that collective properties of
%multiple motor teams can be understood in terms of single molecule
%properties, with the only effective interaction between motors arising
%from load-sharing between like motors. A complete  quantitative
%understanding of the properties of multiple motor teams requires a
%more detailed understanding of how motors interact with each
%other and with the cargo\cite{KUNWAR,NEW2,BOUZAT1,BOUZAT2}. 
%The modification of the mathematical form of dynein detachment
%rate with load, which we arrived at from purely phenomenological
%arguments here, might then be a manifestation of an underlying collective
%emergent property of a team of dynein motors, although recent experimental results\cite{KUNWAR1} suggest that
%it is present at the single-molecule level itself.  If so, 

To summarize, we believe that our modeling efforts have yielded important insights into the mechanisms of bidirectional transport
driven by TOW between asymmetric motor teams. The highlight of the present paper is the deduction of the non-monotonic nature of the detachment rate of dynein as a function of 
opposing load, from stall-time measurements done it {\it in vitro}.  We find it interesting and reassuring that the modified detachment rate proposed fits well with {\it in vitro} stall-time data with beads in optical traps, as well as with
tug-of-war data from endosomes. In light of the recent {\it in vitro} experimental results with dynein from {\it Drosophila}\cite{KUNWAR1}, our studies suggest that this might a universal property of cytoplasmic dynein.

What could be the underlying mechanism for the non-Kramers form of 
the detachment rate? Recently, Driver et. al. \cite{DIEHL} have suggested that load-dependent transport behaviour of 
multiple-motor teams can be understood better using models which explicitly include intermediate
states between the bound and free states, and the transitions between
these. Perhaps, a consideration of such intermediate states could provide a clue to the saturation of the dynein detachment rate beyond stall force. 

From a statistical perspective, the bidirectional motion of cargoes carried by motors of opposing polarity is 
a biased random walk, but with a strong history-dependence. The origin of this history-dependence (in the present model) lies in the effective interaction between opposing motors via load-sharing. We have provided a glimpse of the novel 
features that arise from this interaction in Fig.\ref{fig:fig5} which shows that the probability of direction reversal following a TOW depends on the direction of the preceding run. A more systematic and extensive investigation of the statistical properties of bidirectional cargo motion with varying motor numbers is presently under progress and will be reported elsewhere.

\ack
M.G would like to thank Roop Mallik and other members of the Motor
Proteins Lab, Department of Biological Sciences, TIFR, Mumbai for many
fruitful discussions and sharing of their experimental data at various
stages. MG also thanks RM for a careful reading of the manuscript and many valuable
suggestions, as well as
bringing reference\cite{MCKINNEY} to our attention. Both the authors
acknowledge useful and illuminating discussions with A. Kunwar. We also acknowledge two anonymous referees for 
valuable suggestions in the first round of reviews.

\end{document}